\documentclass[a4paper,11pt]{article}
\pdfoutput=1 
\usepackage{style}  
\usepackage{epstopdf}
\usepackage{subfigure}
\usepackage[sort&compress]{natbib} 
\setcitestyle{square}
\theoremstyle{plain}

\theoremstyle{definition}

\usepackage{mathtools}
\usepackage{enumitem}
\theoremstyle{remark}

\usepackage{verbatim}
\usepackage{blindtext}
\usepackage[T1]{fontenc}
\usepackage{lipsum,cuted}
\usepackage{float}
\usepackage{comment}
\usepackage{graphicx}
\graphicspath{{figures/}} 
\usepackage{xcolor}
\usepackage{siunitx}

\usepackage[bottom, hang]{footmisc}
\usepackage{caption}
\newcommand{\orcid}[1]{\href{https://orcid.org/#1}{\textcolor[HTML]{A6CE39}{\aiOrcid}}}
\usepackage{titlesec}
\titlespacing*{\section}{0pt}{1.1\baselineskip}{\baselineskip}

\title{Limited Angle Tomography reconstruction for non-standard MBI system by means of parallel-hole and pinhole optics}

\author[a,b,1]{G. E. Poma\note{Corresponding author.}}
\author[c]{, F. Garibaldi}
\author[c]{, F. Giuliani}
\author[d]{, T. Insero}
\author[c]{, M. Lucentini}
\author[e]{, A. Marcucci}
\author[f]{, P. Musico}
\author[g]{, J. Nuyts}
\author[c]{, F. Santavenere}
\author[g]{, G. Schramm}
\author[a]{, C. Sutera}
\author[c]{ and E. Cisbani}

\affiliation[a]{INFN - Sezione di Catania, Via S. Sofia 64, 95123, Catania, Italy}
\affiliation[b]{Dipartimento di Fisica ed Astronomia, Universit\`a di Catania, Via S. Sofia 64, 95123, Italy}
\affiliation[c]{ISS, Istituto Superiore di Sanit\`a, Viale Regina Elena 199, 00161, Rome, Italy}
\affiliation[d]{Ospedale Pediatrico ''Bambin Ges\`u'', Viale di San Paolo 15, 00146, Rome, Italy}
\affiliation[e]{Aeronautica Militare Italiana, Viale dell'Universit\`a 4, 00185, Rome, Italy}
\affiliation[f]{INFN - Sezione di Genova, Via Dodecaneso 33, 16146, Genoa, Italy}
\affiliation[g]{KU Leuven - University of Leuven, Department of Imaging and Pathology, Nuclear Medicine \& Molecular imaging, Medical Imaging Research Center (MIRC), B-3000 Leuven, Belgium}

\emailAdd{elio.poma@ct.infn.it}

\abstract
    {
      The purpose of the present work is the study of reconstruction properties of a new Molecular Breast Imaging (MBI) device for the early diagnosis of breast cancer, in Limited Angle Tomography (LAT), by using two asymmetric detector heads with different collimators. The detectors face each other in anti-parallel viewing direction and, mild-compressing the breast phantom, they are able to reconstruct the inner tumour of the phantoms with only a limited number of projections using a dedicated maximum-likelihood expectation maximization (ML-EM) algorithm. Phantoms, MBI system, as well as Monte Carlo simulator using Geant$4$ Application for Tomographic Emission (GATE) software, are briefly described. MBI system's model has been implemented in IDL (Interactive Data Visualization), in order to evaluate the best LAT configuration of the system and its reconstruction ability by varying tumour's size, depth and uptake. LAT setup in real and simulated configurations, as well as the ML-EM method and the preliminary reconstruction results, are discussed.
    }


\begin{document} 
	\maketitle
	\flushbottom
	
	\section{Introduction} 
	\label{sec:intro}
	Breast cancer (BC) is the most common cancer among women, with almost $1.7$ million new cases  occurring among women worldwide in $2012$ and about $30$\% resulting deaths. Every $1$ in $8$ women ($12$\%) in the United States will develop invasive BC during their lifetime. Its successful treatment is related to the diagnosis in early stage, that is the detection of tumours smaller than $1$ cm \cite{bc.org}.
	\\The standard BC screening exam is mammography, thanks to its high sensitivity ($<90$\%) and specificity (almost $90$\%); however it  can be affected by several factors, particularly in dense breast where sensitivity drops down to $30$-$48$\% \cite{elmore} while women with dense breast have higher probability to develop the cancer than the non-dense breast's women.
	\\New types of imaging techniques could overcome these limitations by complementing anatomical with functional imaging: ultrasound (US), magnetic resonance (MRI), positron emission tomography (PET) and single photon emission computed tomography (SPECT).
	\\MRI is not always able to distinguish the difference between cancerous abnormalities, because of the many false positive results, which may lead to unnecessary breast biopsies. US is also a promising adjunct to mammography, particularly for discriminating between benign cysts and malignant tumours: it provides real-time images independently of density, but on the other side many cancers are not visible. 
	\\Nuclear medicine molecular imaging methods like PET and SPECT (in the form of Molecular Breast Imaging, MBI), have also been used for detection and diagnosis of BC. These techniques are not affected by breast density and are powerful complements to morphological imaging modalities, because of their ability to provide functional informations thanks to radio-pharmaceutical uptake in the breast tumours. This can help differentiate malignant from benign tumours.
	\\Single and multiple heads SPECT devices are commercially available, but conventional Anger gamma cameras suffer from a limitation, that is the spatial resolution \cite{peterson}: in fact, due to the large tumour-to-detector distances, they are only able to detect tumours greater than $10$ mm. 
	\\To overcome these limitation, an innovative and portable MBI system with two detector heads (large parallel-hole or LH, and small pinhole or SH) has been developed at "Istituto Superiore di Sanit\`a" in Rome, as second level exam for the small-size BC diagnosis. Thanks to the pinhole optics, the system can spot-compress the breast detecting the small tumours ($\le5$ mm) with increased system sensitivity.
	\\The LH with sensitive area of $150\times200$ mm$^2$ hosts a parallel hole collimator, while the SH ($50\times50$ mm$^2$) has a pinhole collimator \cite{garibaldi, brevetto}: the two detector heads are facing each other in anti-parallel viewing direction, and the pinhole collimator can be focused on a small region in the breast, magnifying the tumour. The MBI system can work also in Limited Angle Tomography (LAT) mode, with the SH pinhole nearby the breast and in rotation around the breast and the LH in fixed position. This configuration produces a single LH projection image and a limited number of SH projection images, acquired while rotating over a finite angle (i.e. $\ang{60}$ of angular coverage).
	\\Our work aims at studying the tomographic reconstruction of tumour as a function of its position, uptake and the SH angular aperture. To this end, a Monte Carlo simulator and an analytic model have been implemented in Geant$4$ Application for Tomography Emission (GATE) and Interactive Data Language (IDL) frameworks, respectively, as well as a dedicated maximum-likelihood expectation maximization (ML-EM) iterative reconstruction method. Different MBI configurations$\backslash$performances have been tested varying LAT key parameters, using the real model's LAT setup in the experimental campaigns \cite{poma_iworid}.
	\\The ultimate aim is to integrate the MBI device with a tomosynthesis device to obtain functional and morphological multi-modality informations: while MBI provides emission images originated from the gamma emitted by the radio-pharmaceutical injected inside the body patient, X-rays are used to generate transmission images of structures inside the body.
	\begin{figure*}[!ht]
		\begin{center} 
			\centering
			\begin{minipage}[ht]{.48\textwidth}
				\centering	
				\includegraphics[width=0.75\textwidth,height=0.88\textheight,keepaspectratio]{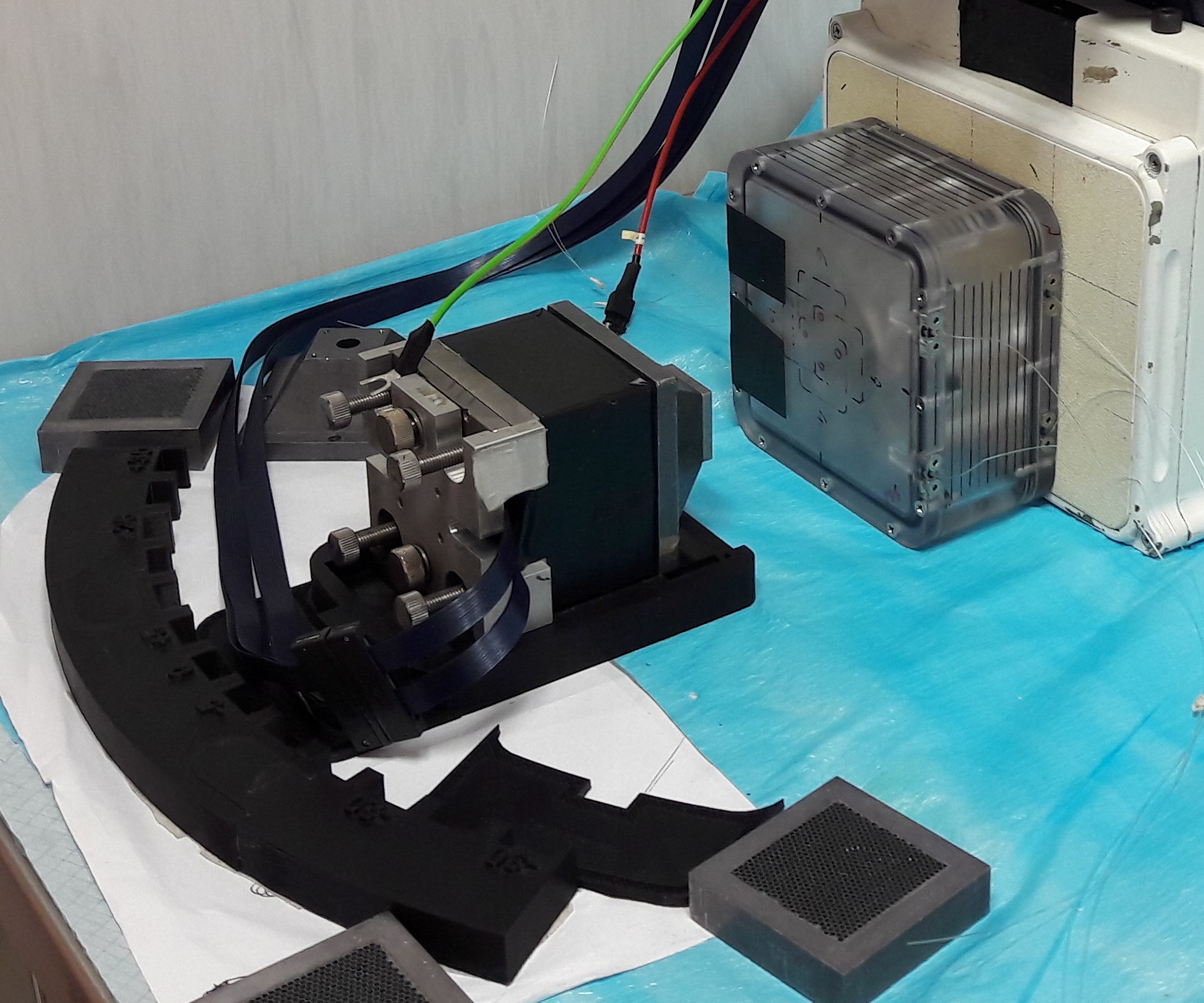}
				\label{lat_real}
			\end{minipage}  
			\begin{minipage}[ht]{.48\textwidth}
				\centering	
				\includegraphics[width=0.75\textwidth,height=0.88\textheight,keepaspectratio]{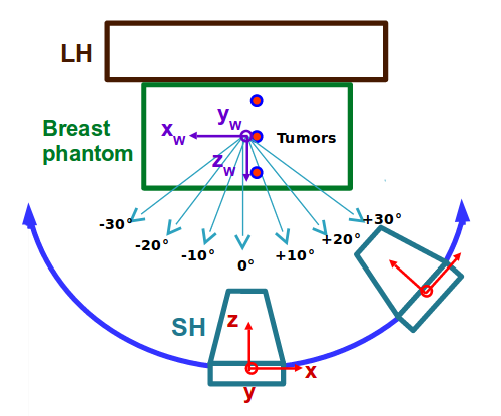}
				\label{lat_real_scheme}
			\end{minipage} 
			\caption{\small Left: side view of MBI device; the small head SH (left) with tungsten housing, supported by 3 axes mechanics and the white painted large head LH (right); the transparent box is the breast phantom. Right: top view of MBI system with the laboratory (purple) and SH (red) reference systems, and three tumours inside breast phantom at different depths.}
			\label{mbi_lat_real}
		\end{center}
	\end{figure*}

	\section{Material and Methods} 
	\label{sec:mbisystem}
	Figure \ref{mbi_lat_real} (left) shows the novel MBI system in LAT configuration, with the two asymmetric detector heads LH and SH in anti-parallel viewing direction.
	\\Thanks to the SH compression functionality, the tumour-to-detector distances can be minimized, so the sensitivity is maximized and the different optics of the two heads contribute to an excellent spatial resolution. The reconstruction of the simultaneous projection images from the two detectors offers an improved Signal-to-Noise Ratio (SNR) and spatial resolution, as well as the tumour localization.
	\\In the baseline setup, the SH consists of an interchangeable tungsten pin-hole collimator (hole diameter range $0.6$-$2.05$ mm, $30\times30\times5$ mm$^3$ tungsten plate) at $4.5$ cm from a pixellated NaI(Tl) scintillator crystal ($50.8\times50.8\times6$ mm$^3$, $1.2$ mm pixel pitch). A parallel hole collimator ($50.8\times50.8\times20.9$ mm$^3$) can replace the pinhole in the SH mainly for test purposes. 
	\\The LH has a bigger active area ($200\times 150$ mm$^2$): it hosts an high sensitivity lead parallel hole collimator ($20.9$ mm length hexagonal septa, $1.475$ mm pixel pitch and $0.305$ mm septal thickness) coupled to an array of pixellated NaI(Tl) ($1.2$ mm pixel pitch, $5$ mm thick). 
	\\Both scintillators are finally coupled to multi-anode PMTs Hamamatsu H$8500$ with $64$ independent channels each \cite{h8500}: a single PMT in the small detector and $4\times3$ PMT array in the large one.
	The $832$ anodes of the $13$ PMTs are readout as independent channels with adjustable gains by a dedicated electronics.
	\\In the current implementation, the device can rotate around a horizontal axis, the distance between the two heads can be manually adjusted, as well as the relative position of the SH with respect to the LH. The small detector can also rotate around a second horizontal axis perpendicular to the above. All these degrees of freedom allow an optimal spot compression, with reasonable comfort of the patient.
	\\The dedicated breast and tumour phantoms have been built to emulate a real compressed breast containing a lesion at different depths. The breast phantom is made of a complex parallelepiped perspex vessel ($120\times120\times64$~mm$^3$, $0.92$ l capacity), which can be filled by water or radioactive liquid. A movable PMMA support inside the breast phantom, can host up to 4 tumours simultaneously: in the current LAT test (figure \ref{mbi_lat_real}, left), only one tumour ($4$ mm radius) was used and imaged at $3$ depths and $7$ rotation angles of SH \cite{poma_iworid}.
	\\In order to optimize the LAT system, evaluate its small tumour imaging ability and improve its performance, given the system complexity, two simulators have been developed in LAT configuration: 
	\begin{itemize}[noitemsep] 
		\item a microscopic processes simulator based on the GATE framework \cite{gate}, using Monte Carlo simulations \cite{poma_cherne} that reproduce the real system, the phantoms and the physics processes described in section \ref{sec:mbisystem};
		\item an analytic model, implemented in IDL software \cite{idl_site}: it emulates the final $3$-D image (different tumour depth and size, breast background activity and other distortion effects) corresponding to the real case (figure \ref{mbi_lat_real}, right), in which tumours are reconstructed by varying the angular coverage, and it uses ray-tracing projectors (RTP) to reconstruct it.
	\end{itemize}
	
	\subsection{GATE model} 
	\label{sec:gate}
\begin{figure*}[!ht]
		\begin{center} 
			\centering
			\begin{minipage}[ht]{.48\textwidth}
				\centering	
				\includegraphics[scale=0.45,keepaspectratio]{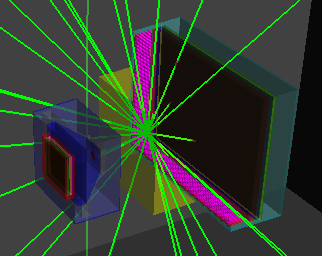}
				\label{lat_gate}
			\end{minipage}  
			\begin{minipage}[ht]{.48\textwidth}
				\centering	
				\includegraphics[scale=0.28,keepaspectratio]{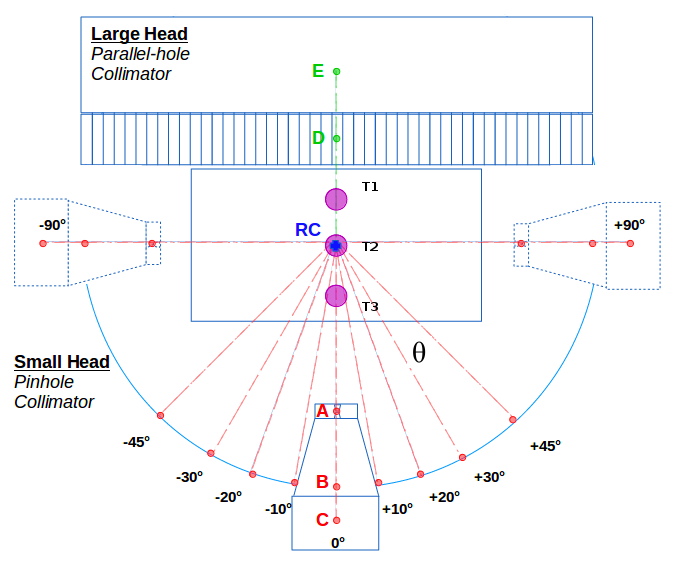}
				\label{lat_gate_scheme}
			\end{minipage} 
			\caption{\small Left: GATE simulator in LAT mode: breast phantom (yellow box) with a tumour inside it emitting $\gamma$-rays from $^{99m}$Tc decay, LH (SH) system to the right (left) of the phantom. Right: MBI sketch in LAT mode, used in GATE and IDL frameworks: rotation center RC, tumour position T, pinhole center A, small head system center B, small head crystal center C, parallel-hole center D, large head crystal center E, small head rotation angle $\theta$.}
			\label{mbi_lat_gate}
		\end{center}
	\end{figure*}

	Figure \ref{mbi_lat_gate} left, reports a picture of the MBI system in LAT configuration as modelled by GATE. The main simulation goals are: the understanding of the effects of different geometrical configurations on the system sensitivity and spatial resolution; the evaluation of the potential performances of the two detector heads; the improvement of the image reconstruction algorithms comparing the simulated images to the acquired ones. 
	\\For these purposes, the implemented LAT model (figure \ref{mbi_lat_gate}, left) is rather flexible and configurable, within the GATE framework: a $8$~mm diameter spherical tumour, placed in the breast phantom origin, emulates a breast cancer lesion that emits $\gamma$-rays, while being acquired by the small head on the left, and the large head on the right. 
	\\The small detector rotates around the breast phantom, pointing to its center, at various angles: a first set of simulations has been done in the angular range $[\ang{-30}, \ang{30}]$ by a step of $\ang{10}$; a second set includes in addition the angles $\pm \ang{45}$, $\pm \ang{90}$; tumour was imaged at each rotation angle of the SH, and different tumour depths.
	\\Right figure \ref{mbi_lat_gate} shows the simulation setup in GATE (breast phantom is a box), which is the same of the analytic model presented in figure \ref{mbi_lat_idl_ideal}, left, but the breast phantom is here a sphere. The $8$ mm diameter-size tumour was simulated at three different depths: in the phantom origin, shifted $20$ mm toward the LH and then $20$ mm toward the SH.
	\\The raw customized output of the GATE simulation consists of the information related to the origin and interactions of the radiation in the sensitive volumes (scintillator and detector). It also holds the optical photons information, their generation and tracking in the scintillation crystals. 
	\\A shell script automates the data simulation and digitization, using the informations on the detector and collimator type, the source diameter and its position inside the breast phantom, the number of $\gamma$-rays emitted isotropically from the source, and the $3$-D position of the small head in LAT configuration. Preliminary results are presented in the following.
	
	\subsection{RTP model}
	\label{sec:idl}
	In order to create and then validate a reliable reconstruction method, a simplified and analytic MBI system's model has been created in IDL, using dedicated routines, developed for the tomographic reconstruction of PET and SPECT devices, used for clinic applications \cite{louvre}.
	\begin{figure*}[!ht]
		\begin{center} 
			\centering
			\begin{minipage}[ht]{.45\textwidth}
				\centering	
				\includegraphics[scale=0.33,keepaspectratio]{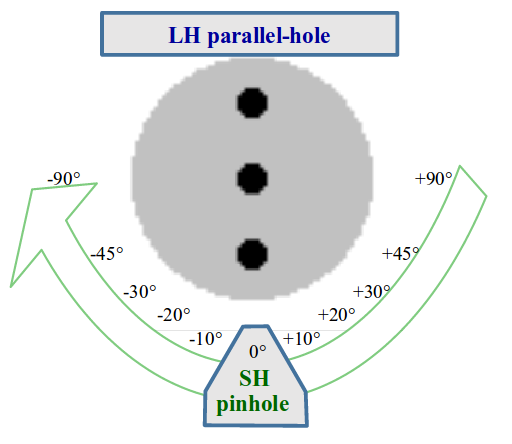}
			\end{minipage}  
			\begin{minipage}[ht]{.45\textwidth}
				\centering	
				\includegraphics[scale=0.28,keepaspectratio]{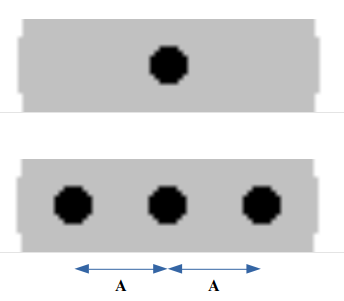}
			\end{minipage}
			\caption{\small Left: breast and three tumours phantoms implemented in IDL, with a sketch of SH rotation around the breast phantom's origin (LH and phantoms are fixed). For simplicity, breast phantom was built as a sphere with a diameter equals to the small yellow box's axis ($64$ mm). Right: sagittal and coronal views of breast phantom, looking at its center which coincides with the position of central tumour. Parameter A is the mutual distance between two close tumours.} 
			\label{mbi_lat_idl_ideal}
		\end{center}
	\end{figure*}
	\\The GATE model explained in section \ref{sec:gate} has been translated in IDL environment and characterized versus the MBI and LAT parameters:
	\begin{itemize}[noitemsep]
		\item large head components: 
		\begin{itemize}[noitemsep]
			\item parallel-hole collimator: $20.5$ mm septal length, $1.475$ mm hole width, $0.305$ mm septal thickness
			\item distance between detector and image center: $59$ mm
			\item intrinsic detector resolution: $1.5$ mm
		\end{itemize}
		\item small head components: 
		\begin{itemize}[noitemsep]
			\item pinhole collimator: $2.05$ mm diameter of pinhole;
			\item distance between detector and image center: $108.5$ mm
			\item intrinsic detector resolution: $1.5$ mm
		\end{itemize}
		\item phantoms:
		\begin{itemize}[noitemsep]
			\item breast: sphere of $32$ mm radius, with attenuation of $140$ keV $\gamma$-rays and background activity $A_{bkg}$
			\item tumours: spheres of R radius at three different depths, tumour-to-tumour distance $A=20$ mm (figure \ref{mbi_lat_idl_ideal}, right), activity $A_{tum}=6.12\times10^7$ such that tumour uptake $U_T = A_{tum}/A_{bkg}$
		\end{itemize}
		\item rotation of small head: $\theta$ rotation angle
		\item reconstructed image: $80\times80\times20$ mm$^3$, with $1$ mm pixel size
	\end{itemize}
	The ray-tracing projector accounted for the position dependent resolution of the parallel-hole and pinhole collimators and the position dependent sensitivity of the pinhole collimator. The tomographic reconstruction of the simulated setup, implemented in IDL, was performed by means of the ML-EM iterative method \cite{mlelm}: $100$ iterations were used for the phantom data reconstruction at each tumour depth.
	\\To suppress limited angle artefacts, the likelihood was combined with a total variation prior (extending the ML approach to a maximum a posteriori (MAP) approach).
	
	\vspace{0.5cm}
	\section{Tomographic reconstruction} 
	\label{sect:tr}
	The first step of MBI simulator's characterization is the tomographic reconstruction of simplified IDL model. Two kinds of simulations have been run, as a function of two different tumour dimensions and uptake, and the small head pinhole's angular coverage: the first one by means of only the pinhole projections, the second one using both detectors projections. The aim of the simulations is the determination of the minimum pinhole angular coverage for the optimal reconstruction of a small tumour, and at the same time understanding if the small head is sufficient or not to localize the tumour activity.
	\\After that, the second step is the test of GATE projection images obtained by previous simulations using the same ML-EM method.

	\subsection{RTP projections} 
	\label{sec:idl_images}
\begin{figure}[!ht]
		\centering
		\includegraphics[width=.8\textwidth, height=0.4\textheight, keepaspectratio]{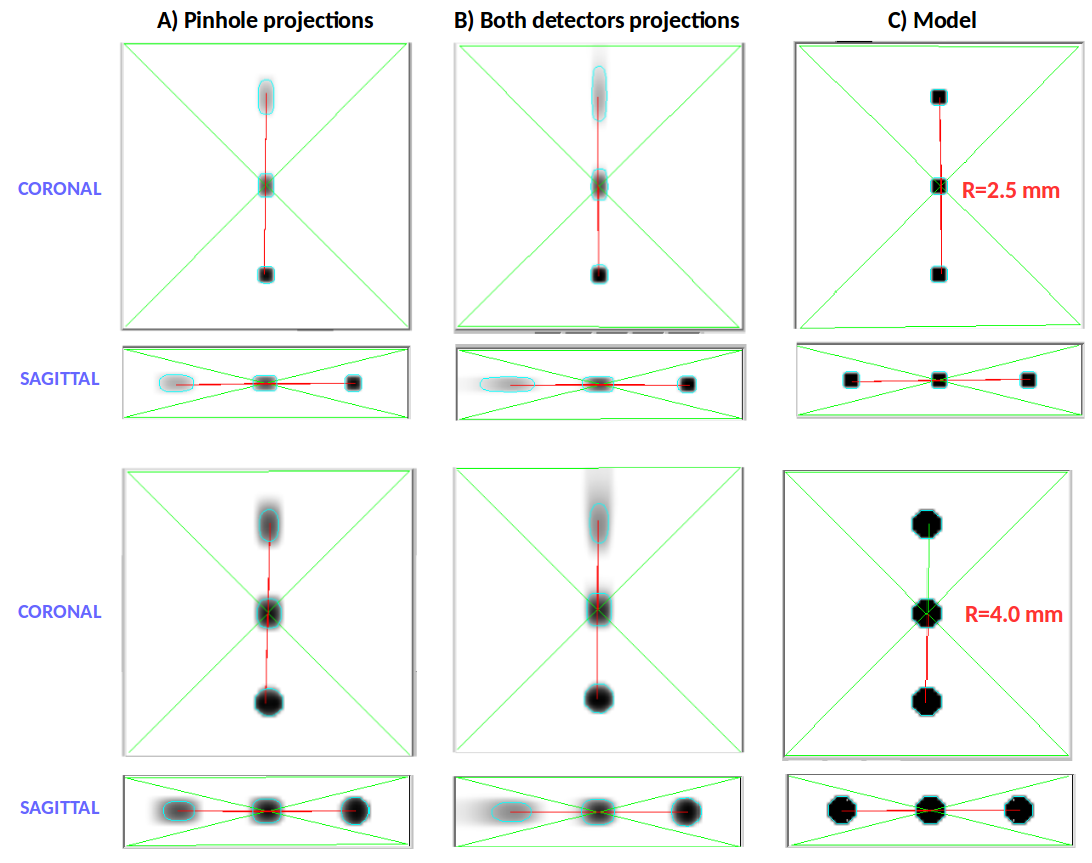} 
		\caption{\small 3-D reconstructions (A and B) of the corresponding analytic model (C), in coronal and sagittal views: tumour radius $R$, breast background $A_{bkg}=0$, $\theta \in [\ang{-30}, \ang{30}]$ by $\ang{10}$ step.} 
		\label{r11_r12_r21_r22}
	\end{figure}

\begin{figure}[!ht]
		\centering
		\includegraphics[width=.8\textwidth, height=0.4\textheight, keepaspectratio]{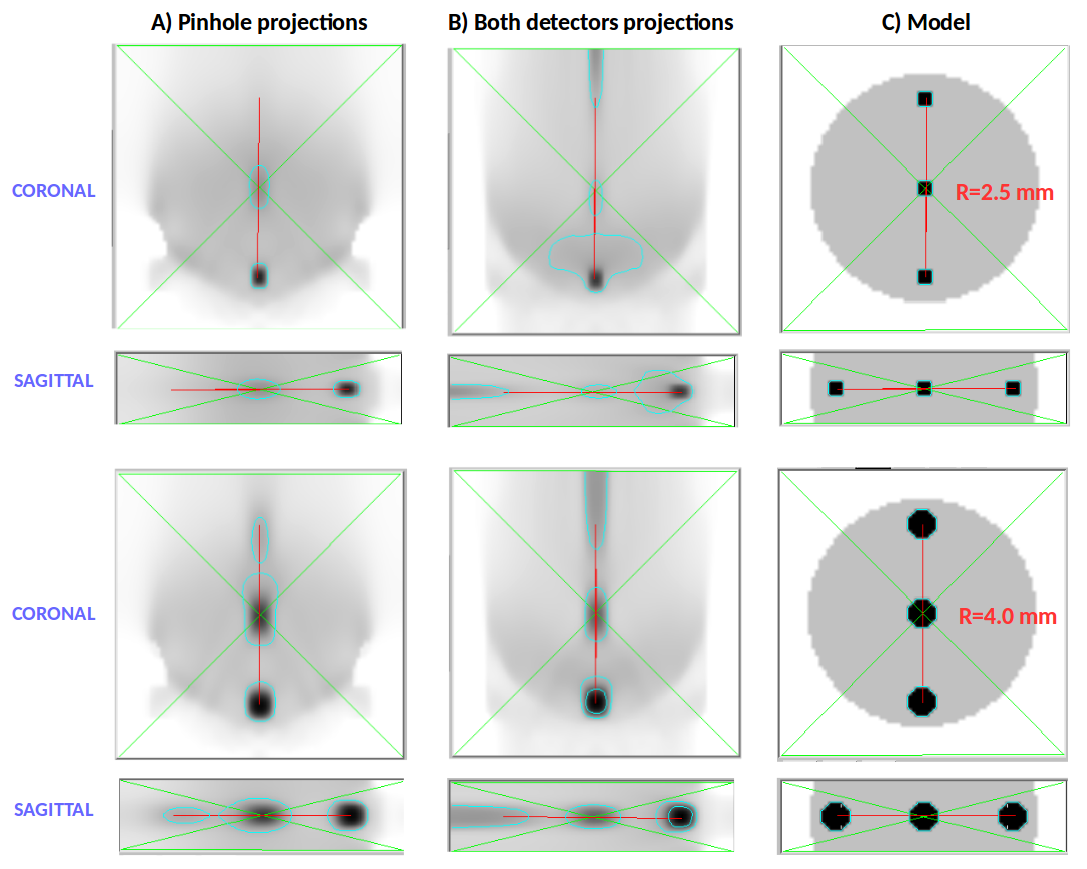} 
		\caption{\small 3-D reconstructions (A and B) of the corresponding analytic model (C), in coronal and sagittal views: tumour radius $R$, tumour uptake $U_T = 10$, $\theta \in [\ang{-30}, \ang{30}]$ by $\ang{10}$ step.}
		\label{r31_r32_r41_r42}
	\end{figure}

\begin{figure}[!ht]
		\centering
		\includegraphics[width=.8\textwidth, height=0.4\textheight, keepaspectratio]{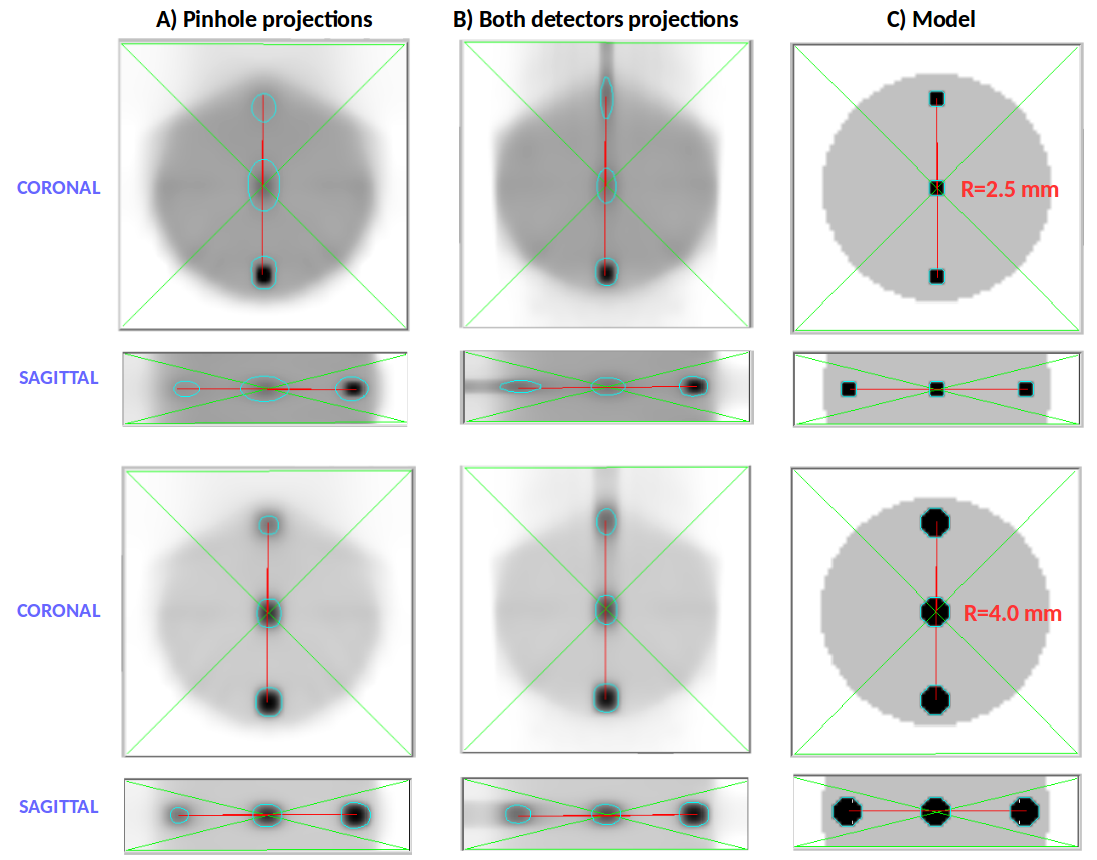} 
		\caption{\small 3-D reconstructions (A and B) of the corresponding analytic model (C), in coronal and sagittal views: tumour radius $R$, tumour uptake $U_T = 10$, rotation angle \\$\theta \in [\ang{-90},\ang{-45},\ang{-30},\ang{-20},\ang{-10},\ang{0},\ang{10},\ang{20},\ang{30},\ang{45},\ang{90}]$.}
		\label{r51_r52_r61_r62}
	\end{figure}

	The simulated system faithfully represents the real one (figure \ref{mbi_lat_real}, right), except for the simplified spherical breast phantom and for the $2.5$ or $4$ mm radius of tumours, chosen to evaluate the spatial resolution limit. The small spheres are fixed in three different positions ($A=25~$ mm, figure \ref{mbi_lat_idl_ideal} right) and the small head always points to the phantom center: each spherical tumour emits $6.12\times10^6$, $140.2$ keV, $\gamma$-rays which are attenuated by the water medium, with a tumour-to-breast activity concentration ratios $U_T$.
	\\The ML-EM algorithm was used to reconstruct an image from the $7$ or $11$ pinhole and the single parallel hole projections. Figures \ref{r11_r12_r21_r22}, \ref{r31_r32_r41_r42}, \ref{r51_r52_r61_r62} show the ML-EM reconstruction using only pinhole projector (A) or both projections (B), in coronal (up) and sagittal (down) slices: each image represents the tumours reconstruction at different depths inside the breast phantom, using or not the breast background $A_{bkg}$, and changing the tumour size.
	\\The blue iso-contours around the tumours (figures \ref{r11_r12_r21_r22}, \ref{r31_r32_r41_r42}, \ref{r51_r52_r61_r62}) define the activity concentration's levels (in \% of maximum value): one blue line corresponds to $1$ level, two lines to $2$ levels. For each level, a \% of maximum value has been chosen: small contours mean small \% levels ($\le 50$\%), while large ones mean $>50$-$60$\%.
	~
	~
	\\The green crosses have been drawn to fix the image center: it coincides with the central tumour. The red line represents the distance between the two farthest tumours: its length is $2A = 50$ mm.
	\\When tumour activity only is present (figure \ref{r11_r12_r21_r22}), each tumour is well reconstructed in both configurations (A) and (B). This is less true when breast activity background is added: using a tumour uptake $U_T = 10$ (figure \ref{r31_r32_r41_r42}) and $\theta \in [\ang{-30}, \ang{30}]$, tumour's reconstruction and depth localization improves in case (B). Finally, if more projections are acquired, tumour identification improves remarkably even for the small tumours (figure \ref{r51_r52_r61_r62}). 
	\\Reconstruction suffers also from the blurring in the direction perpendicular to the parallel hole detector. Tumours are imaged with better resolution when they were closer to the pinhole collimator, due to the the pinhole's higher sensitivity and resolution. This is also true because the projection lines have a wider angular range than when the lesion is farther away: a wider angular range reduces the limited angle artefacts.

	\subsection{GATE projections} 
	\label{sec:gate_images}
	The above reconstruction method has been extended to the MBI GATE model described above (section \ref{sec:gate}).
	One tumour ($R = 4$ mm), emitting $6.12\times10^7$ $\gamma$-rays in $240$ seconds ($A_{tum} = 255$ kBq) is located at three different depths inside the breast phantom (figure \ref{mbi_lat_gate}, right): T$2$ coincides with the breast phantom center, while T$1$ and T$3$ are $20$ mm away from it, and close to large and small head, respectively. In the preliminary study, no breast background activity was added ($A_{bkg} = 0$).
	\begin{figure}[!ht]
		\centering
		\includegraphics[width=.8\textwidth, height=0.4\textheight, keepaspectratio]{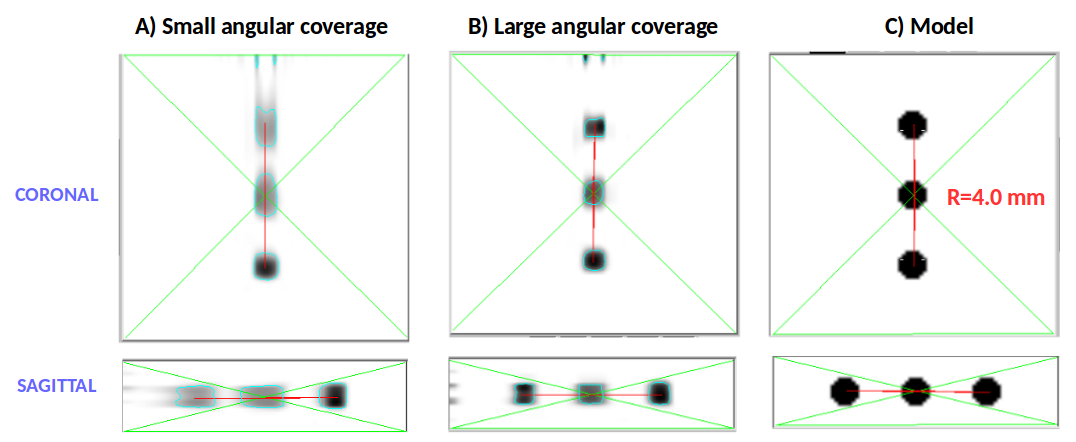} 
		\caption{\small ML-EM reconstruction using GATE projections of large and small heads: tumour $R=4.0$ mm, $A_{bkg} = 0$. A): $\theta \in [\ang{-30},\ang{30}]$; \\B): $\theta \in [\ang{-90},\ang{-45},\ang{-30},\ang{-20},\ang{-10},\ang{0},\ang{10},\ang{20},\ang{30},\ang{45},\ang{90},]$; C): reference model.}
		\label{r30_r90}
	\end{figure}
	\\As in the previous analytic analysis, tumour identification and its depth localization improve with the angular coverage (figure \ref{r30_r90}). GATE simulations of MBI system, in LAT configuration, are also consistent with the real data acquired in the experimental campaigns, identifying the tumour in function of its depth using only a small angular coverage \cite{poma_iworid}.
	\\Reconstruction from GATE simulations and simulation framework's validation, in a more realistic condition ($A_{bkg}$ and Compton scattering, added), are ongoing.
	
	\vspace{1cm}
	\section{Conclusions} 
	\label{sec:concl}
	All the above analyses have assumed an high sensitivity MBI system, with administered activities of $100-200$ MBq $^{99m}$Tc-sestamibi. This results in a breast dose of $\approx 0.25$ mGy and an effective dose of $\approx 1.1$ mSv to the body; the corresponding doses of digital mammography are $4$ mGy (mean glandular dose to the breast) and $0.5$ mSv (effective dose to the patient) \cite{oconnor}.
	The MBI system with small and large heads, demonstrated the ability of detection and reconstruction of tumours smaller than $5$ mm in Limited Angle Tomography configuration, improving the standard gamma camera's spatial resolution and SNR. Ray-tracing projection and GATE models and simulations have been implemented and performed, respectively, and an ML-EM reconstruction algorithm with total variation regularization has been used for the tumour's identification and its depth localisation inside a breast phantom. Preliminary reconstruction images suggests that the system is able to identify small tumours by increasing the angular coverage of small head pinhole, and by using both detectors projections.
	Improvements on Monte Carlo simulations and ML-EM method are ongoing.

\end{document}